%% file: paper.tex
\documentclass{sig-alternate-05-2015}

\usepackage[latin1]{inputenc}
\usepackage{paralist} 
\usepackage[colorinlistoftodos,disable]{todonotes} 
\usepackage{multirow} 
\usepackage{subfigure} 
\usepackage{ifdraft}
\usepackage[square,numbers]{natbib}
\usepackage[nonumberlist,acronym]{glossaries}
\usepackage{xspace}
\usepackage{xcolor,colortbl}
\usepackage{hhline}
\usepackage{footmisc}

\usepackage{hyperref}
\hypersetup{
    hidelinks = true,
}

\newcommand{\jesus}[1]{\todo[inline,color=blue!60]{\footnotesize{Jesus: #1}}}

\newacronym{hipaa}{HIPAA}{Health Insurance Portability and Accountability Act}
\newcommand{\hipaa}{\Gls{hipaa}\xspace}
\newacronym{paas}{PaaS}{Platform as a Service}
\newcommand{\paas}{\Gls{paas}\xspace}
\newacronym{byod}{BYOD}{Bring Your Own Device}

\newacronym{dsl}{DSL}{Domain Specific Language}

\newcommand{\dsls}{\Glspl{dsl}\xspace}
\newacronym{sla}{SLA}{Service Level Agrement}

\newcommand{\slas}{\Glspl{sla}\xspace}
\newacronym{pci}{PCI-DSS}{Payment Card Industry Data Security Standard}
\newcommand{\pci}{\Gls{pci}\xspace}
\newacronym{soa}{SOA}{Service Oriented Architecture}
\newcommand{\soa}{\Gls{soa}\xspace}
\newacronym{bpm}{BPM}{Business Process Management}
\newcommand{\bpm}{\Gls{bpm}\xspace}
\newacronym{cisa}{CISA}{Cybersecurity Information Sharing Act}
\newcommand{\cisa}{\Gls{cisa}\xspace}
\newacronym{gdpr}{GDPR}{General Data Protection Regulation}
\newcommand{\gdpr}{\Gls{gdpr}\xspace}

\begin{document}


\CopyrightYear{2016} 
\setcopyright{acmlicensed}
\conferenceinfo{SEAMS'16,}{May 16-17 2016, Austin, TX, USA}
\isbn{978-1-4503-4187-5/16/05}\acmPrice{\$15.00}
\doi{http://dx.doi.org/10.1145/2897053.2897070}

%

\title{Towards Adaptive Compliance}
%
%
%
%
%

\numberofauthors{1}
\author{
\alignauthor
Jes\'{u}s Garc\'{i}a-Gal\'{a}n$^{1}$,
Liliana Pasquale$^{1}$, 
George Grispos$^{1}$,
Bashar Nuseibeh$^{1,2}$\\
      \vspace{1.2mm}
      \affaddr{$^1$ Lero - The Irish Software Research Centre, University of Limerick, Ireland}\\
      \vspace{0.8mm}
      \affaddr{$^2$ Department of Computing \& Communications, The Open University, Milton Keynes, UK}\\
      \vspace{0.5mm}
}



\maketitle
\begin{abstract}
  Mission critical software is often required to comply with multiple
  regulations, standards or policies. Recent paradigms, such as cloud
  computing, also require software to operate in heterogeneous, highly
  distributed, and changing environments. In these environments, compliance
  requirements can vary at runtime and traditional compliance
  management techniques, which are normally applied at design time, may no longer be sufficient.  
  In this paper,
  we motivate the need for \emph{adaptive compliance} by illustrating possible compliance concerns determined by runtime variability. We further motivate our work by means of a cloud
  computing scenario, and present two main contributions.
  First, we propose and justify a process to support adaptive compliance that
  extends the traditional compliance management lifecycle with the activities 
  of the Monitor-Analyse-Plan-Execute (MAPE)
  loop, and enacts adaptation through re-configuration.  Second,
  we explore the literature on software compliance and classify existing
  work in terms of the activities and concerns of adaptive compliance.
  In this way, we determine how the literature can support our proposal
  and what are the open research challenges that need
  to be addressed in order to fully support adaptive compliance.
\end{abstract}

%
%


%
%

%
%



\input{intro}


\input{background}

\input{motivation}

\input{solution}

\input{runtimeSolution}

\input{models}
\input{challenges}

\input{discussion}

\section*{Acknowledgements}

We acknowledge SFI grant 10/CE/I1855 and ERC Advanced Grant no. 291652 (ASAP).
We also thank Mark Mcgloin for later discussions on software compliance.


\setlength{\bibsep}{1.72pt plus 0.3ex}

\footnotesize

\newpage
\bibliography{bibliography}

\end{document}

%% file: intro.tex
\section{Introduction}
\label{sec:intro}

With software becoming increasingly pervasive, ensuring compliance to regulations, standards or policies is also becoming increasingly important to foster its wider adoption and acceptability by society and business. For example, in recent years, compliance with industrial regulations (e.g., \hipaa) and data security standards (e.g., \pci and ISO/IEC 27000-series) has become an essential requirement of some software systems. Non-compliance can result in loss of reputation, financial fines\footnote{\url{http://www.hhs.gov/about/news/2014/05/07/data-breach-results-48-million-hipaa-settlements.html}} or even criminal prosecution. 
Within acade\-mia, compliance has been examined in the areas of requirements engineering~\cite{otto2007addressing}, \soa~\cite{Tilburg08}, cloud computing~\cite{martens2011risk} and \bpm~\cite{fellman2014state}. 
Each of these has tackled compliance from different perspectives, including the interpretation of regulations into compliance requirements~\cite{breaux2008analyzing, ghanavati2014goal}, compliance checking~\cite{awad2008efficient, muller2014comprehensive}, and the definition of a reference process for compliance management~\cite{turetken2012capturing,martens2011risk}.

Ensuring compliance is more challenging in software systems that operate in heterogeneous, highly distri\-buted and changing environments, such as cloud computing services.
Cloud providers often deliver their services to clients from different geographical locations that have their own compliance requirements. Providing customised \emph{Compliance-as-a-Service} could relieve clients of the compliance burden and give providers a significant competitive advantage. However, cloud providers may still face different and multi-jurisdictional compliance requirements. They must also comply with the regulations that apply where their physical infrastructure resides. In a multi-tenancy environment, in which different clients may share computational resources, this could also lead to overlaps and conflicts between different compliance requirements. 
All this variability may in turn lead to compliance violations. Although compliance at runtime has gained attention recently \cite{birukou2010integrated, gomez2015compliance, ly2015compliance}, as far as we are aware, existing techniques are normally applied at design time and are unable to deal with this kind of runtime variability.

In this paper, we propose \emph{adaptive compliance} as \emph{the capability of a software system to continue to satisfy its compliance requirements, even when runtime variability occurs}. We motivate our work by using a \paas scenario and provide two main contributions. 
First, we propose and justify a process to support adaptive compliance that extends the traditional compliance management lifecycle with the activities of the Monitor-Analyse-Plan-Execute (MAPE) loop, and achieves adaptation through reconfiguration.
Second, we explore the literature on software compliance to identify which activities of our process are already supported and which present open research challenges. 
Our ambition is to motivate the need of adaptive compliance and encourage researchers from the adaptive systems community to address these challenges.

The rest of the paper is organised as follows. Section~\ref{sec:background} introduces the main concepts and terminology adopted in software compliance. Section~\ref{sec:motivation} presents a motivating scenario that illustrates the concerns when handling runtime variability in compliance. Section~\ref{sec:solution} describes our adaptive compliance process and its existing support in the literature.  Section~\ref{sec:challenges} describes research challenges related to adaptive compliance. Finally, Section~\ref{sec:discussion} concludes the paper.


%% file: background.tex
\begin{figure}[t!!]
\centering
\includegraphics[scale=0.48]{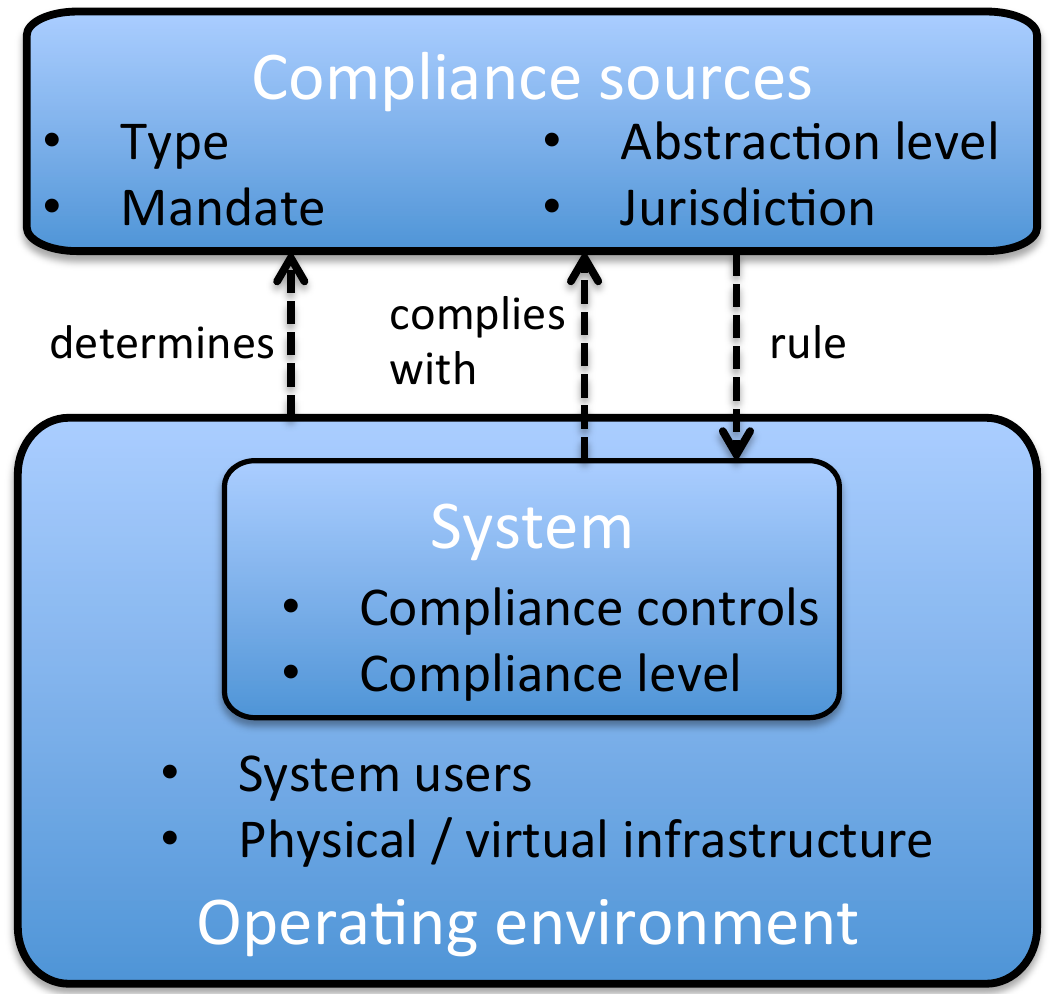} 
\caption{Compliance elements and dimensions.}
\label{fig:dimensions}
\end{figure}

\section{Compliance in Software}
\label{sec:background}

In the context of information systems, compliance refers to ``ensuring that an organisation's software and system conform with multiple laws, regulations and policies'' \cite{Zdun12IEEE}. From this definition we can distinguish two main elements in compliance: the \emph{system} and the \emph{compliance sources} that have to be conformed with. Moreover, the system runs in an \emph{operating environment}, that may affect the compliance sources against which the system has to conform. 
Figure~\ref{fig:dimensions} shows these elements and their different dimensions.

\footnotetext[2]{\url{http://www.hhs.gov/hipaa/}}
\footnotetext[3]{\url{https://www.congress.gov/bill/114th-congress/senate-bill/754}}

The \emph{type} of compliance sources refers to the kind of rules specified in the source. In particular, a compliance source can include regulations such as \hipaa\footnotemark[2], \cisa\footnotemark[3], and \gdpr\footnotemark[4]; standards such as \pci\footnotemark[5] and ISO/IEC 27000 series\footnotemark[6]; good practices; and internal policies within a particular organisation. \emph{Mandate} refers to the optional or mandatory character of the compliance source. For example, regulations are compulsory (e.g., \hipaa in the US) while standards and internal policies (e.g., ISO/IEC 27000 series) may not. The \emph{abstraction level} denotes the level of interpretation necessary to enact the statements mandated by a compliance source in a system. Regulations are usually expressed at a higher level of abstraction than standards and internal policies. For example, \gdpr requires companies to prove compliance without suggesting specific mechanisms, while an internal organisation policy may specifically state that rooted Android phones cannot connect to the company's internal network. Compliance sources apply to particular \emph{jurisdictions}, territories or spheres of activity. For example, \hipaa applies to US organisations dealing with personal healthcare information, while the \gdpr is intended to apply to any organisation delivering services to EU citizens.

A system executes \emph{compliance controls}, which are implementations of the rules defined by a compliance source. For example, section 164.312(a)(2)(iii) within \hipaa mandates the implementation of procedures to log off from an electronic session after a predetermined time of inactivity. A compliance control to address this rule could involve forcing user sessions to expire after five minutes of inactivity. Compliance controls determine the \emph{compliance level} of a system with regards to its compliance sources. Three \emph{de-facto} levels of compliance are proposed in the literature \cite{colombo2014business}: `full compliance' when all rules are satisfied; `partial compliance' when all mandatory rules are satisfied; and `non-compliance' where one or more mandatory rules are not satisfied. 
 
Although the operating environment is highly dependent on the specific application domain, we distinguish two main characterising elements. First, the \emph{system users} who can reside and/or operate in different geographical locations and/or spheres of activity may need to comply with different compliance sources. Second, the \emph{physical or virtual infrastructure} where a system operates influences the applicable compliance sources and the compliance level that needs to be achieved. For example, IT systems in a US hospital have to comply with \hipaa and doctors who use personal devices to access patient records become part of the operating environment. 

\begin{figure}[t!!]
\centering
\includegraphics[scale=0.5]{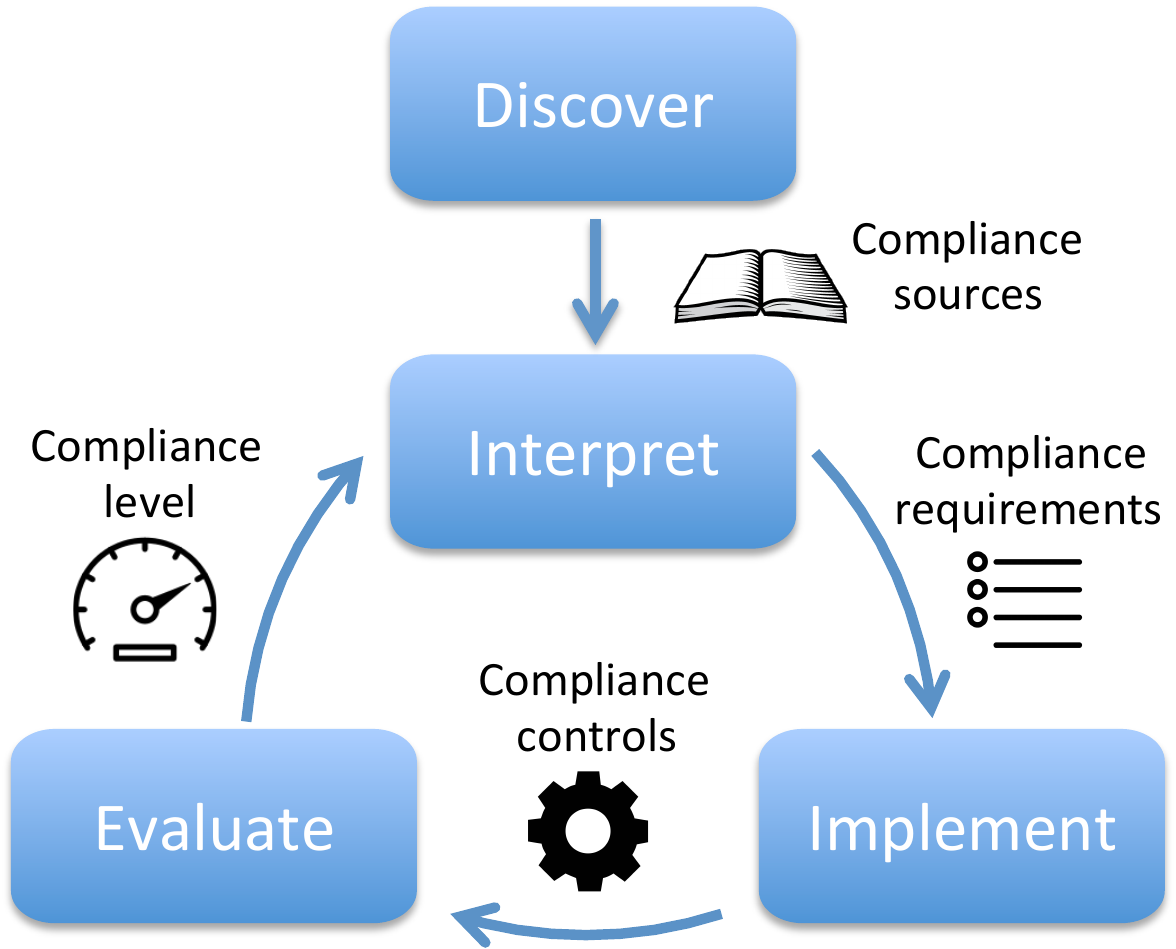} 
\caption{Compliance management lifecycle.}
\label{fig:ComplianceArtefacts}
\end{figure}

\footnotetext[4]{\url{http://ec.europa.eu/justice/data-protection/}}
\footnotetext[5]{\url{https://www.pcisecuritystandards.org}}
\footnotetext[6]{\url{http://www.iso.org/iso/catalogue\_detail?csnumber=66435}}

\jesus{should we highlight the shortcomings of the existing lifecycle approaches?}

Different reference processes have been proposed in the literature to support compliance management ~\cite{cabanillas2011exploring, daniel2009business, ramezani2014supporting, Tilburg08}. Figure~\ref{fig:ComplianceArtefacts} presents a simplified compliance management lifecycle covering the main activities of these processes. Initially, \emph{compliance sources} are discovered depending on the system and its operating environment. Compliance sources are then interpreted to extract \emph{compliance requirements}, 
which are expressed as rules about actors and their rights and obligations on particular data objects~\cite{breaux2008analyzing}. 
During development, the requirements are implemented in the system as \emph{compliance controls}. 
Finally, compliance requirements and controls are evaluated to determine the compliance level they ensure and to assess how they can be improved, if necessary.


%% file: motivation.tex
\section{Motivating Scenario}
\label{sec:motivation}

In this section we present a \paas scenario, shown schematically in Figure~\ref{fig:PaaSScenario}, to motivate adaptive compliance and illustrate the main compliance concerns arising from runtime variability. The \paas provider offers customers a technological stack offering Database Management Systems (DBMS), run-time environments and various frameworks. The stack is deployed on top of an infrastructure which is hosted in the United States. Customers can deploy their own software applications using the stack offered by the \paas provider. 

The \paas provider also aims to provide \emph{compliance-as-a-service} to its clients, that is the capability to satisfy on-demand the compliance needs of its clients. As shown in Figure~\ref{fig:PaaSScenario}, initially the provider has to satisfy the compliance requirements of two different clients (Client 1 and 2). Client 1 operates in the US and stores patient health records, which are accessed by other third-party organisations. Hence, Client 1 has to comply with \hipaa regulations. Client 2 is also located in the US and handles its client's credit card information using the \paas. Therefore, Client 2 has to comply with \pci. Client 1 and Client 2 have differing compliance requirements, which the \paas provider must ensure are satisfied.


Although some cloud providers guarantee compliance for particular
regulations (e.g., HIPAA compliance by Catalyze\footnotemark[7]\footnotetext[7]{\url{https://catalyze.io/}} and
TrueVault\footnotemark[8]\footnotetext[8]{\url{https://www.truevault.com/}}), they are usually unprepared to satisfy emergent differing and varying compliance requirements. 
New \paas clients could introduce new compliance demands that need to be traded-off against those of existing clients who share the same execution platform. In our scenario, Client 3 is a new client providing financial services in the US and therefore requires a certifiable degree of privacy and security (e.g., ISO/IEC 27018). 
Clients can also change their compliance requirements due to changes in the jurisdictions that apply to their services. For example, Client 3 is considering expanding its operations to Europe and therefore it will need to comply with the European Union's \gdpr, which requires explicit user authorisation for any re-purposing of their personal data. This could result in a direct conflict with \cisa, which authorises sharing of personal data with federal institutions. 

Furthermore, variability in the compliance sources and the operating environment can also affect the compliance requirements and their satisfaction. 
In the case of compliance sources, Client 1, for example, could be required to comply with \cisa in the near future. 
While compliance sources may rarely evolve, changes in the system and its operating environment can occur more frequently. 
For example, updates to the DBMS may affect how data encryption is supported and hence, the satisfaction of the compliance requirements.
This also includes changes in the physical infrastructure of the service. 
In this sense, the \paas provider could move some of its data centres to Europe, which would trigger the need to comply with EU regulations for data retention and management.
The variability exposed by this scenario can be summarised by the following main concerns:

\begin{figure}[t!]
\centering
\includegraphics[scale=0.42]{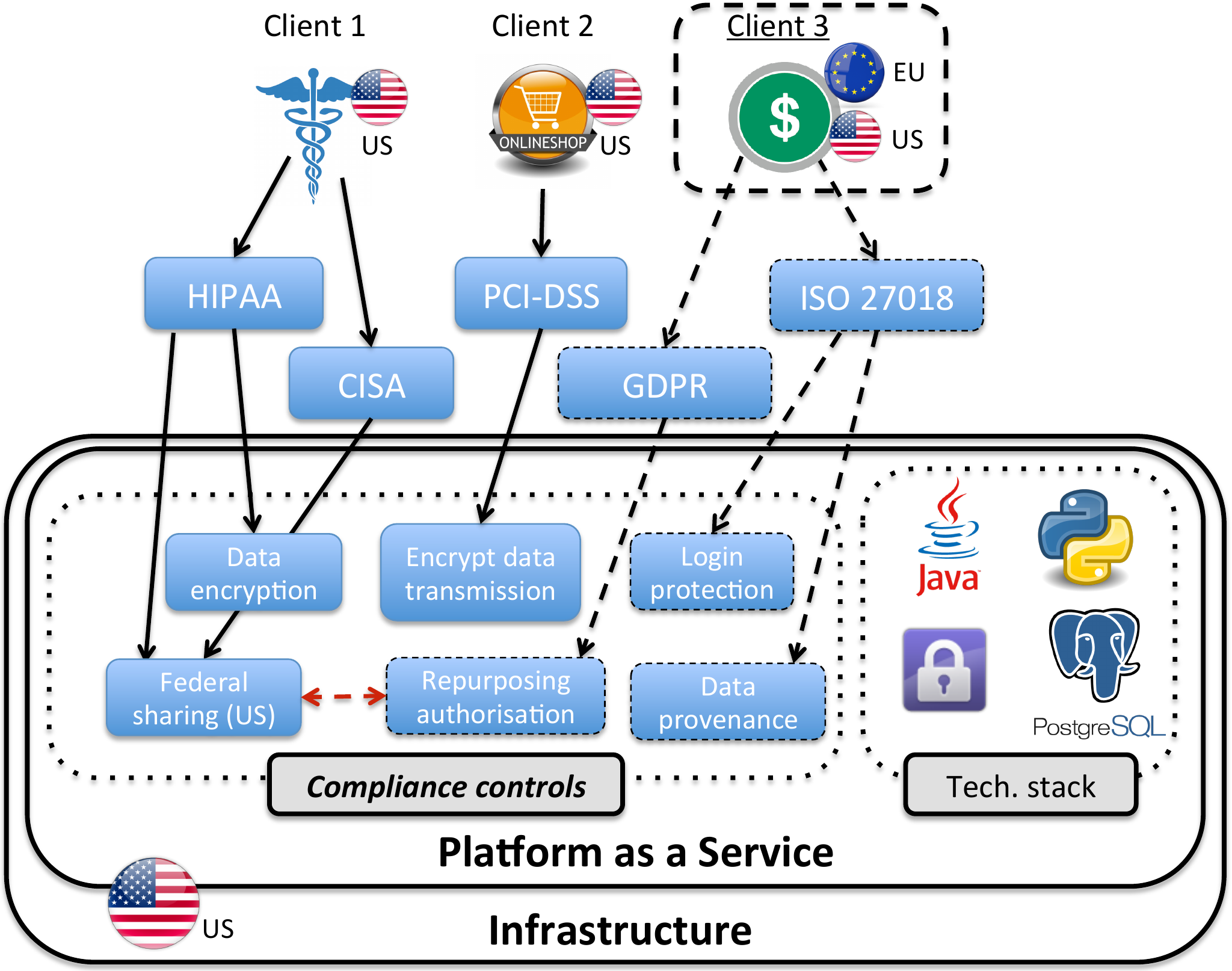} 
\caption{PaaS Scenario.}
\label{fig:PaaSScenario}
\end{figure}

%

\begin{inparaenum}[\itshape a\upshape)]
\item \textbf{Awareness}. Runtime variability requires awareness of any changes that take place in the operating environment, the system and the compliance sources which can impact compliance satisfaction. In particular, \paas clients must be able to elicit and modify their preferences with respect to the compliance sources they need to satisfy. The provider also needs to be aware of infrastructure changes and assess how these changes impact the compliance requirements.

\item \textbf{Automation}: Compliance-as-a-service requires executing appropriate compliance controls using a dynamic approach. This would also involve automating the discovery and interpretation of compliance sources, as well as the identification and remedy of compliance violations and potential conflicts between compliance requirements. 

\item \textbf{Assurance}: The service provider needs to produce assurances about whether or not the compliance level is that required by its clients. These assurances can be provided by collecting data, showing traceability between compliance controls and requirements or by delivering formal proofs.

\item \textbf{Performance}: The on-demand nature of cloud computing means that a cloud provider has to respond to changes in a timely manner so as to avoid service outages.
\end{inparaenum}

\jesus{connect these concerns with the next section}

%% file: solution.tex
\begin{figure*}[ht!!]
\centering
\includegraphics[scale=0.7]{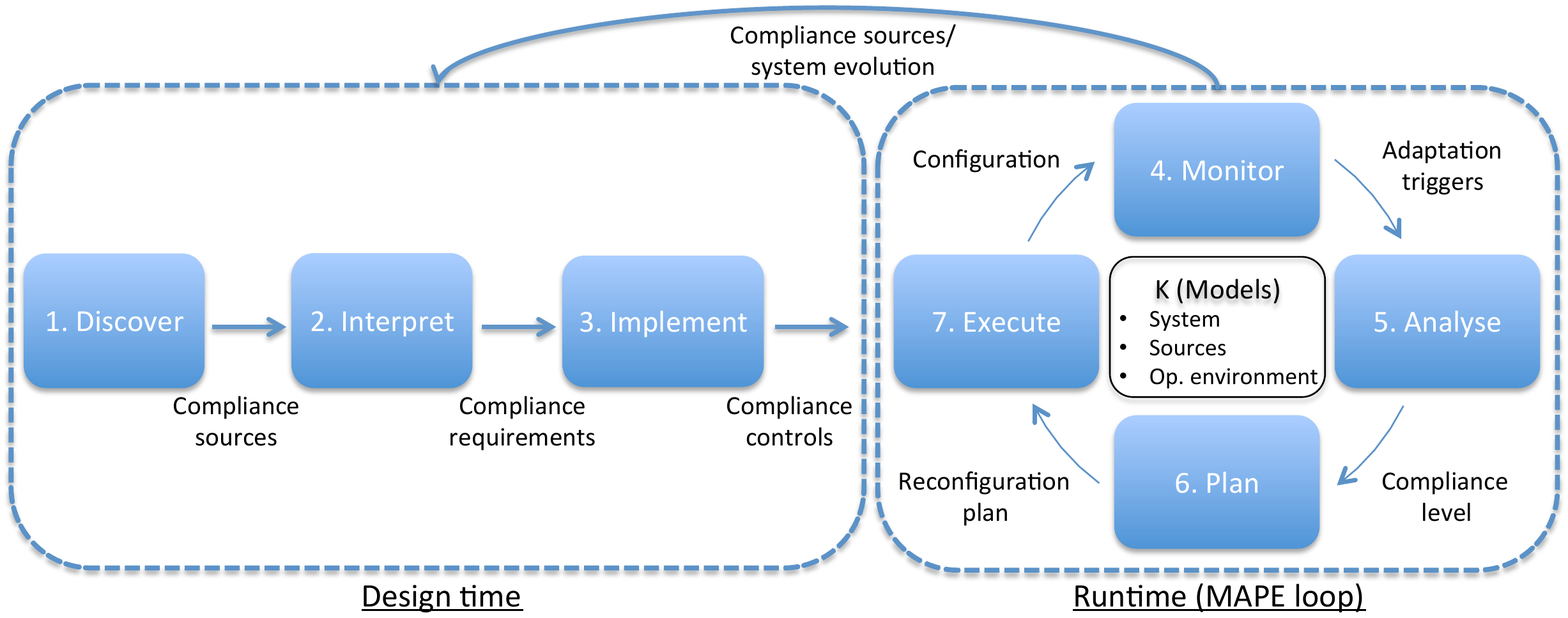} 
\caption{Adaptive compliance process.}
\label{fig:AdaptiveCompliance}
\vspace{-10pt}
\end{figure*}

\section{Adaptive Compliance}
\label{sec:solution}

\jesus{read the beginning of this section again}

\jesus{discuss the differences with the traditional MAPE loop and the limitations (human-in-the-loop?)}

In this section we present a process to achieve adaptive compliance and a summary of the support provided by the existing literature. This process, shown in Figure~\ref{fig:AdaptiveCompliance}, extends the compliance management lifecycle of Figure~\ref{fig:ComplianceArtefacts} with added support for variability runtime concerns through the MAPE loop. 
The adaptive compliance process begins with the automated \emph{discovery} of compliance sources with which the system needs to comply. Factors that influence the discovery of compliance sources include the system's physical location, its sphere of activity and the potential stakeholders. 
Next, the compliance sources must be \emph{interpreted} in order to identify the compliance requirements, which demand close collaboration between legal and domain experts, and software engineers~\cite{otto2007addressing}.
This activity could benefit from mechanisms to share and reuse compliance requirements, such as a multi-organisation repository. 
The requirements are subsequently \emph{implemented} in the system as compliance controls. Since the system is intended to meet differing compliance needs at runtime, the compliance controls should be flexible enough to be enabled, disabled and customised when required. 

At runtime, the system must \emph{monitor} its own state, the operating environment and the compliance sources. Any changes in these elements may result in either compliance violations or a reduced compliance level. While changes in compliance sources take place slowly, changes in the system or the operating environment often require a response at runtime. 
When changes are detected, the system must \emph{analyse} their impact on the compliance level. First, overlaps  between the applying compliance requirements should be analysed, since they might lead to conflicts and consequently compliance violations. Second, the compliance level must be checked and if compliance violations are found, these must be diagnosed to determine their causes. 
In that case, the system needs to \emph{plan} a reconfiguration of the compliance controls to improve the compliance level when possible. 
Finally, the computed reconfiguration must be \emph{executed} in the system, effectively improving the compliance level.

This process requires of ``live'' models, especially at runtime to enact the Knowledge (K) component of the MAPE loop. 
Such models must describe the compliance sources and its requirements, the system and its compliance controls, and also the operating environment, including user preferences and the system infrastructure. 
The relevance of these models depends on the particular activities.
Some of them are more important at design time (e.g., compliance sources for their discovery and interpretation), while others are necessary at runtime (e.g., operating environment for the monitoring, or compliance controls for the plan). 

In the following, we explore how the compliance literature supports the adaptive compliance activities and addresses the concerns presented in Section~\ref{sec:motivation}. This analysis allows us to identify topics which have been well discussed, along with gaps leading to research challenges. Table~\ref{tab:survey} relates existing approaches supporting the activities with the concerns that these approaches have addressed. Partially addressed areas are shown in light grey, while areas not addressed are shown in dark grey. 

\begin{table*}[ht!!]
\centering
\begin{tabular}{|l r|p{1.6cm}|p{1.6cm}|p{1.8cm}|p{1.7cm}|p{1.8cm}|p{1.6cm}|p{1.6cm}|}
\cline{3-9}
\multicolumn{2}{c}{} & \multicolumn{7}{|c|}{\textbf{Adaptive Compliance Activities}} \\
\cline{3-9}
\multicolumn{1}{c}{} & & \multicolumn{1}{c|}{\textbf{Discover}} & \multicolumn{1}{c|}{\textbf{Interpret}} & \multicolumn{1}{c|}{\textbf{Implement}} & \multicolumn{1}{c|}{\textbf{Monitor}} & \multicolumn{1}{c|}{\textbf{Analyse}} & \multicolumn{1}{c|}{\textbf{Plan}} & \multicolumn{1}{c|}{\textbf{Execute}} \\
\hline
\multirow{5}{*}[-6ex]{\rotatebox{90}{\textbf{Concerns}}}
& \multicolumn{1}{|r|}{ \textbf{Awareness}} & \cellcolor{gray!30} \cite{kerrigan2003logic,otto2007addressing,Tilburg08} & \cite{boella2013managing, breaux2008analyzing, Ghanavati:2015:ILI:2821464.2821473, ghanavati2014goal, gordon2012reconciling, maxwell2012managing, schmidt2012assessing, siena2013automated} & \cellcolor{gray!30} \cite{koetter2014integrating, ramezani2014supporting, schleicher2009maintaining} & \cellcolor{gray!30}
\cite{birukou2010integrated, ly2015compliance, muller2014comprehensive, van2012context} & \cellcolor{gray!30}
\cite{ghanavati2014goal, gordon2012reconciling, Knuplesch15ER, van2012context}
 & \cellcolor{gray!30} \cite{cabanillas2011exploring, ghose2007auditing, ly2012enabling} & N/A \\
\hhline{|~ |-|-|-|-|-|-|-|-|} 
& \multicolumn{1}{|r|}{\textbf{Automation}} & \cellcolor{gray!30}
\cite{boella2013managing,kerrigan2003logic}&
\cellcolor{gray!30} \cite{panesar2013supporting, toval2002legal, tran2012compliance, turetken2012capturing} &  \cellcolor{gray!30} \cite{ramezani2014supporting, SchummLMS10, tran2012compliance} & \cite{birukou2010integrated, muller2014comprehensive, van2012context} & \cite{awad2008efficient, birukou2010integrated, elgammal2014formalizing, gomez2015compliance, kerrigan2003logic, molina2012model, muller2014comprehensive, ramezani2012did, siena2013automated, van2012context} & \cellcolor{gray!100} \cite{ghose2007auditing} & \cellcolor{gray!100} \\
\hhline{|~ |-|-|-|-|-|-|-|-|} 
& \multicolumn{1}{|r|}{\textbf{Assurances}} & N/A &
\cite{boella2013managing, breaux2008analyzing, Ghanavati:2015:ILI:2821464.2821473, siena2013automated, turetken2012capturing}
& \cite{panesar2013supporting, tran2012compliance} & N/A & \cite{gomez2015compliance, muller2014comprehensive, ramezani2012did} & \cellcolor{gray!100} & \cellcolor{gray!100} \\ 
\hhline{|~ |-|-|-|-|-|-|-|-|} 
& \multicolumn{1}{|r|}{\textbf{Performance}} & N/A & N/A & \cellcolor{gray!100} & N/A & \cellcolor{gray!100} \cite{colombo2014business} & \cellcolor{gray!100} & \cellcolor{gray!100} \\
\hhline{|- |-|-|-|-|-|-|-|-|} 
\end{tabular}
\caption{Overview of the literature, structured by the concerns and activities of adaptive compliance.}
\label{tab:survey}
\vspace{-10pt}
\end{table*}

\subsection{Discover}

\jesus{what do we mean with each concern? -> awareness and automation}
Although discovery is a fundamental activity in the compliance management lifecycle, it has received little attention from the research community. 
Some studies have highlighted its importance~\cite{otto2007addressing, Tilburg08}, but without defining factors that affect applying compliance sources. 
Some studies have provided repository tools \cite{boella2013managing, kerrigan2003logic}. \citet{kerrigan2003logic} describe environmental regulations by means of an XML-based format and facilitate the discovery through searchable concept hierarchies. \citet{boella2013managing} provide the Eunomos web-based system to manage knowledge about laws and legal concepts in the financial sector.
However, in general, discovery lacks automated support and a general taxonomy of factors.

\subsection{Interpret}

The interpretation of compliance sources has been widely covered by the research literature~\cite{otto2007addressing}.
However, most of the existing work relies on partially or totally manual techniques to extract compliance requirements. 
These techniques include Semantic Parameterisation~\cite{breaux2008analyzing}, goal based analysis, and CPR (commitment, privilege and right) analysis~\cite{schmidt2012assessing}, which have been validated by means of empirical studies.
Some authors have focused on compliance requirements variability, and in particular on the multiple possible interpretations of a regulation~\cite{boella2013managing, Ghanavati:2015:ILI:2821464.2821473, siena2013automated} and the evolution of the requirements~\cite{maxwell2012managing}. 
The analysis and reconciliation of potentially conflicting mul\-ti-jurisdictional requirements have also received attention~\cite{gordon2012reconciling, ghanavati2014goal}.
In terms of automation, some research efforts have focused on the description of compliance requirements by using different approaches, such as \dsls \cite{tran2012compliance}, UML \cite{panesar2013supporting} or semi-formal representations \cite{turetken2012capturing}. A repository of compliance requirements has also been proposed~\cite{toval2002legal}, although without real tool support.
Assurances demonstrating the correctness of compliance requirements with respect to the sources have been suggested, especially in terms of traceability links \cite{breaux2008analyzing, Ghanavati:2015:ILI:2821464.2821473}  and formal proofs \cite{turetken2012capturing, boella2013managing, siena2013automated}.

\subsection{Implement}
\label{subsec:Implement}

Compliance implementation has received attention and partial automated support, in particular from the \bpm community.
Several works have considered configurable compliance controls for business processes, in the form of compliance descriptors~\cite{koetter2014integrating}, business process templates~\cite{schleicher2009maintaining} and configurable compliance rules~\cite{ramezani2014supporting}.
Implementation automation has been addressed from different perspectives. While some approaches have proposed an automated derivation of compliance controls from the requirement descriptions~\cite{tran2012compliance}, others have presented repositories of reusable process fragments~\cite{SchummLMS10} or compliance rules~\cite{ramezani2014supporting}.
Some of these also provide support for implementation assurances by explicitly linking compliance controls and requirements~\cite{tran2012compliance}, and concepts of the compliance source to the application domain~\cite{panesar2013supporting}.
However, the impact of compliance controls on the system performance has been surprisingly neglected.

%% file: runtimeSolution.tex
\subsection{Monitor}

The literature on monitoring is mainly focused on system changes, neglecting the compliance sources and the operating environment. While compliance sources rarely change at runtime, the operating environment does, requiring a timely detection and response.
Several works have shown awareness of different monitoring factors, such as the system execution~\cite{birukou2010integrated}, the Quality of Service in \slas~\cite{muller2014comprehensive}, or time, resources and data in business processes~\cite{ly2015compliance}. However, the operating environment has only been considered for particular aspects of specific cases in the context of business processes~\cite{van2012context}.
Most of those works present approaches to automate the monitoring in business processes \cite{birukou2010integrated, van2012context} and \soa~\cite{muller2014comprehensive}.

\subsection{Analyse}

Compliance analysis is the activity that most has attracted most attention from the research community, especially for checking compliance levels. 
However, additional awareness on the potential conflicts of multi-jurisdictional requirements \cite{ghanavati2014goal, gordon2012reconciling} or the impact of the operating environment \cite{Knuplesch15ER, van2012context} is necessary.
Compliance checking can take place at design time and at runtime. 
Design time checking approaches are common for business processes, and rely on a plethora of analytical techniques, such as those based on Petri Nets \cite{awad2008efficient, ramezani2012did} and temporal logic \cite{awad2008efficient, elgammal2014formalizing}. 
Similar approaches have been proposed for general regulatory compliance, by means of inference engines \cite{siena2013automated} and first order predicate calculus \cite{kerrigan2003logic}, and for business to business interactions \cite{molina2012model}.
Runtime compliance checking has gained momentum recently, especially in business processes \cite{birukou2010integrated, gomez2015compliance, van2012context} and \soa \cite{birukou2010integrated, muller2014comprehensive}.
Some of these compliance checking approaches also provide diagnosis support (i.e. assurances) for the causes of compliance violations \cite{gomez2015compliance, muller2014comprehensive, ramezani2012did}.
In general, checking automation and diagnosis are well covered at design and runtime. However, automated multi-jurisdictional analysis and the consideration of the operating environment for the checking remain as open challenges.
Moreover, the performance of all these approaches have been generally overlooked, although compliance checking has been proven to be np-complete for business processes \cite{colombo2014business}.

\subsection{Plan}

To the best of our knowledge, the compliance literature presents very few works supporting the remedy of compliance violations. 
Although some authors have discussed the concept of compliance improvement \cite{ghose2007auditing, cabanillas2011exploring, ly2012enabling}, there is a lack of automated support. 
\citet{cabanillas2011exploring} state the need to provide recovery capabilities when compliance violations are detected at runtime.
\citet{ly2012enabling} discuss the notion of healable compliance violations, i.e. violations that can be fixed by restructuring the process or inserting additional branches, while \citet{ghose2007auditing} present a partially automated technique, based on structural and semantic patterns, to modify non-compliant processes in order to restore compliance. 
Our vision to enact compliance improvement is through the reconfiguration of the compliance controls. In this sense, several authors have proposed configuration capabilities for implementing compliance, as shown in Section~\ref{subsec:Implement}. However, none of these present specific techniques to remedy violations when detected.

\subsection{Execute}


The literature has not paid attention to this activity beyond works on configuration capabilities for compliance.
In our view, the key concerns of this activity should be the automated execution, in a timely manner, of reconfigurations that improve the compliance level. Assurances about how the reconfiguration execution has improved the compliance level should also be provided.

%% file: models.tex
\subsection{Models}
\label{sec:models}

Most of the research on compliance modelling has focused on compliance sources and requirements.
Multiple approaches have been proposed to represent regulations, using XML notations \cite{kerrigan2003logic}, UML  \cite{panesar2013supporting} or particular \dsls \cite{siena2013automated}.
Other works describe additional compliance sources, such \slas extending the WS-Agreement notation \cite{muller2014comprehensive}, or privacy policies using OWL-DL \cite{kahmer2008automating}. 
There are also numerous proposals to represent compliance requirements, especially by means of formal or semi-formal languages \cite{breaux2008analyzing, ghanavati2014goal, gordon2012reconciling} and \dsls \cite{birukou2010integrated, tran2012compliance}.
Nonetheless, the description of compliance controls and the operating environment appear to have been overlooked: while only few works on business processes address the former \cite{koetter2014integrating, SchummLMS10} by means of compliance descriptors and process fragments, the latter is even more neglected \cite{van2012context}, as we have presented in the previous sections.

%% file: challenges.tex
\section{Research Challenges}
\label{sec:challenges}

Adaptive compliance poses a number of research challenges, some carried over from ``traditional'' compliance work, as well as some significant new ones. Open research challenges relevant to adaptive compliance but carried over from previous compliance research include:

\begin{inparaenum}[\itshape 1\upshape)]
\item \textbf{Compliance sources interpretation.} Multiple authors have proposed specific techniques to interpret regulations and extract compliance requirements, especially in requirements engineering. However, existing work is still limited, 
since this activity usually relies on the specific domain knowledge of the requirements engineers and is performed manually.

\item \textbf{Multi-jurisdictional requirements.} The analysis of overlaps between different regulations has also attracted attention, although existing approaches are only partially automated. In order to detect conflicts among different compliance sources at runtime, adaptive compliance requires more fully automated analysis than provided by current techniques. 

\item \textbf{Remedy for compliance violations.} Our work also highlighted the ongoing challenge of automated remedies for compliance violations. Although there are multiple proposals for automated compliance checking and diagnosis, existing mitigation techniques for violations usually require human intervention.
Since the adaptive compliance process has to handle violations at runtime, we need to provide the process with automated, dynamic mitigation approaches.
\end{inparaenum}

Our work also suggests that adaptive compliance raises new research challenges. We identify five main gaps in the literature that must be addressed:

\begin{inparaenum}[\itshape 1\upshape)]
\setcounter{enumi}{3}
\item \textbf{Compliance readiness.} 
We define compliance readiness as the capability of a system to foresee and comply with different compliance requirements. 
This capability requires awareness of different compliance sources and requirements that may apply to the system or its clients, and a pool of compliance controls, ready to be customised and invoked.
Although some authors have envisaged variability in business process rules to respond to variable compliance requirements, those approaches need to be extended for more complex controls and scenarios.

\item \textbf{Compliance automation.} Currently, only compliance checking has been somewhat automated, and even so, 
often overlooking the impact of the operating environment.
Compliance sources discovery and interpretation remain to be automated, although there are some promising partial successes \cite{kerrigan2003logic, boella2013managing, toval2002legal}.
A repository of compliance sources and their different context dependent interpretations, could support this tedious and error-prone activity. Moreover, automating the reconfiguration  of compliance controls in order to correct compliance violations is needed.

\item \textbf{Runtime assurances.} Demonstrating compliance is often just as important as actually being compliant. There are several approaches to demonstrate compliance by tracing compliance requirements to regulations and compliance controls. However, existing support at runtime is more patchy, and focuses on compliance violations diagnosis, primarily considering systems but not their compliance sources nor their operating environment. Therefore, we need to extend the diagnosis to these elements, and provide evidence about if and how reconfigurations really increase compliance levels.

\item \textbf{Models for compliance controls and operating environment.} Our work shows multiple proposals to describe compliance rules and requirements, but very few to describe compliance controls and the operating environment. While the operating environment is highly dependent on the specific application domain, we think that a standard way to describe the compliance controls, their variability, and their impact on the system is necessary.

\item \textbf{Impact on performance.} Surprisingly, existing research has overlooked the effects of compliance on system performance. Since the adaptive compliance process is intended to handle compliance issues at runtime, performance is increasingly important. Therefore, more efficient ways are needed to check compliance, which has already been shown to be an NP-complete problem. Furthermore, empirical studies are needed to assess the impact on the system performance of executing compliance controls and monitoring the operating environment.
\end{inparaenum}

%% file: discussion.tex
\section{Conclusions and Future Work}
\label{sec:discussion}

In this paper, we have attempted to broaden the definition of compliance-as-a-service to include our idea for adaptive compliance. We have proposed a process to achieve adaptive compliance and discussed how existing work can support the various activities of our adaptive compliance process. A short review of the literature has identified that while existing approaches focus on design-time compliance, very little work has examined the increasing run-time variability found in compliance sources, systems and their operational environment. Furthermore, our review was used to identify several future research challenges which need to be addressed in order to fulfil our vision for adaptive compliance.

Although one of our main ambitions is to automate as much of the compliance process as is feasible, there are limits to this. 
Currently, we are better equipped to automate checking and enforcement of compliance rules.
However, neither all our proposed activities nor compliance rules can be fully automated. 
For example, a rule that specifies the behaviour of a security incident response team may need to be crafted manually, and its enforcement depends on enforcing human behaviour. 
This requires a discussion on the ``human-in-the-loop'' aspects of adaptive compliance. 


Our future work will include conducting a more in-depth analysis of the literature to further investigate the compliance process gaps identified in this paper. The objective of this in-depth literature review would be to expand our understanding of related areas such as reconfiguration planning and execution in adaptive systems. As one of our main ambitions is to automate as much of the compliance process as possible, future work will need to examine which parts of the process can be automated and to what extent. This work will also look to address issues with automation through runtime re-configuration. Finally, future work will examine how our idea of adaptive compliance can be extended into other domains, such as the Internet of Things, where a myriad of heterogeneous and potentially untrusted devices interact. That could provide an additional and important cyber-physical perspective to adaptive compliance.

%% file: paper.bbl
\begin{thebibliography}{38}
\providecommand{\natexlab}[1]{#1}
\providecommand{\url}[1]{\texttt{#1}}
\expandafter\ifx\csname urlstyle\endcsname\relax
  \providecommand{\doi}[1]{doi: #1}\else
  \providecommand{\doi}{doi: \begingroup \urlstyle{rm}\Url}\fi

\bibitem[Awad et~al.(2008)Awad, Decker, and Weske]{awad2008efficient}
A.~Awad, G.~Decker, and M.~Weske.
\newblock Efficient compliance checking using bpmn-q and temporal logic.
\newblock In \emph{6th Int. Conf. on Business Process Management}, pages
  326--341, 2008.

\bibitem[Birukou et~al.(2010)Birukou, D'Andrea, Leymann, Serafinski, Silveira,
  Strauch, and Tluczek]{birukou2010integrated}
A.~Birukou, V.~D'Andrea, F.~Leymann, J.~Serafinski, P.~Silveira, S.~Strauch,
  and M.~Tluczek.
\newblock {An Integrated Solution for Runtime Compliance Governance in SOA}.
\newblock In \emph{ICSOC'10}, pages 122--136, 2010.

\bibitem[Boella et~al.(2013)Boella, Janssen, Hulstijn, Humphreys, and van~der
  Torre]{boella2013managing}
G.~Boella, M.~Janssen, J.~Hulstijn, L.~Humphreys, and L.~van~der Torre.
\newblock Managing legal interpretation in regulatory compliance.
\newblock In \emph{14th Int. Conf. on Artificial Intelligence and Law}, pages
  23--32, 2013.

\bibitem[Breaux et~al.(2008)Breaux, Ant{\'o}n, et~al.]{breaux2008analyzing}
T.~D. Breaux, A.~Ant{\'o}n, et~al.
\newblock Analyzing regulatory rules for privacy and security requirements.
\newblock \emph{IEEE Trans. on Software Engineering}, 34\penalty0 (1):\penalty0
  5--20, 2008.

\bibitem[Cabanillas et~al.(2011)Cabanillas, Resinas, and
  Ruiz-Cort{\'e}s]{cabanillas2011exploring}
C.~Cabanillas, M.~Resinas, and A.~Ruiz-Cort{\'e}s.
\newblock {Exploring Features of a Full-coverage Integrated Solution for
  Business Process Compliance}.
\newblock In \emph{Advanced Information Systems Engineering Workshops}, pages
  218--227, 2011.

\bibitem[Daniel et~al.(2009)Daniel, Casati, D'Andrea, Mulo, Zdun, Dustdar,
  Strauch, Schumm, Leymann, Sebahi, et~al.]{daniel2009business}
F.~Daniel, F.~Casati, V.~D'Andrea, E.~Mulo, U.~Zdun, S.~Dustdar, S.~Strauch,
  D.~Schumm, F.~Leymann, S.~Sebahi, et~al.
\newblock {Business Compliance Governance in Service-oriented Architectures}.
\newblock In \emph{Int. Conf. on Advanced Information Networking and Apps.},
  pages 113--120, 2009.

\bibitem[Elgammal et~al.(2014)Elgammal, Turetken, van~den Heuvel, and
  Papazoglou]{elgammal2014formalizing}
A.~Elgammal, O.~Turetken, W.-J. van~den Heuvel, and M.~Papazoglou.
\newblock {Formalizing and Appling Compliance Patterns for Business Process
  Compliance}.
\newblock \emph{Software \& Systems Modeling}, pages 1--28, 2014.

\bibitem[Fellman and Zasada(2014)]{fellman2014state}
M.~Fellman and A.~Zasada.
\newblock {State-of-the-Art of Business Process Compliance Approaches: A
  Survey}.
\newblock In \emph{22nd European Conf. on Information Systems}, 2014.

\bibitem[Ghanavati and Hulstijn(2015)]{Ghanavati:2015:ILI:2821464.2821473}
S.~Ghanavati and J.~Hulstijn.
\newblock {Impact of Legal Interpretation on Business Process Compliance}.
\newblock In \emph{TELERISE}, pages 26--31, 2015.

\bibitem[Ghanavati et~al.(2014)Ghanavati, Rifaut, Dubois, and
  Amyot]{ghanavati2014goal}
S.~Ghanavati, A.~Rifaut, E.~Dubois, and D.~Amyot.
\newblock {Goal-oriented Compliance with Multiple Regulations}.
\newblock In \emph{22nd Int. Conf. Requirements Engineering}, 2014.

\bibitem[Ghose and Koliadis(2007)]{ghose2007auditing}
A.~Ghose and G.~Koliadis.
\newblock Auditing business process compliance.
\newblock \emph{ICSOC'07}, pages 169--180, 2007.

\bibitem[G{\'o}mez-L{\'o}pez et~al.(2015)G{\'o}mez-L{\'o}pez, Gasca, and
  P{\'e}rez-Alvarez]{gomez2015compliance}
M.~T. G{\'o}mez-L{\'o}pez, R.~M. Gasca, and J.~M. P{\'e}rez-Alvarez.
\newblock {Compliance Validation and Diagnosis of Business Data Constraints in
  Business Processes at Runtime}.
\newblock \emph{Information Systems}, 48:\penalty0 26--43, 2015.

\bibitem[Gordon and Breaux(2012)]{gordon2012reconciling}
D.~G. Gordon and T.~D. Breaux.
\newblock Reconciling multi-jurisdictional legal requirements: a case study in
  requirements water marking.
\newblock In \emph{20th Int. Conf. on Requirements Engineering}, pages 91--100,
  2012.

\bibitem[Kahmer et~al.(2008)Kahmer, Gilliot, and Muller]{kahmer2008automating}
M.~Kahmer, M.~Gilliot, and G.~Muller.
\newblock Automating privacy compliance with expdt.
\newblock In \emph{ECE/CEC}, pages 87--94. IEEE, 2008.

\bibitem[Kerrigan and Law(2003)]{kerrigan2003logic}
S.~Kerrigan and K.~H. Law.
\newblock {Logic-based Regulation Compliance-assistance}.
\newblock In \emph{9th Int. Conf. on Artificial Intelligence and Law}, pages
  126--135, 2003.

\bibitem[Knuplesch et~al.(2015)Knuplesch, Fdhila, Reichert, and
  Rinderle-Ma]{Knuplesch15ER}
D.~Knuplesch, W.~Fdhila, M.~Reichert, and S.~Rinderle-Ma.
\newblock {Detecting the Effects of Changes on the Compliance of
  Cross-organizational Business Processes}.
\newblock In \emph{34th Int. Conf. on Conceptual Modeling}, 2015.

\bibitem[Koetter et~al.(2014)Koetter, Kochanowski, Weisbecker, Fehling, and
  Leymann]{koetter2014integrating}
F.~Koetter, M.~Kochanowski, A.~Weisbecker, C.~Fehling, and F.~Leymann.
\newblock {Integrating Compliance Requirements across Business and IT}.
\newblock In \emph{EDOC}, pages 218--225, 2014.

\bibitem[Ly et~al.(2012)Ly, Rinderle-Ma, G{\"o}ser, and Dadam]{ly2012enabling}
L.~T. Ly, S.~Rinderle-Ma, K.~G{\"o}ser, and P.~Dadam.
\newblock {On Enabling Integrated Process Compliance with Semantic Constraints
  in Process Management Systems}.
\newblock \emph{Information Systems Frontiers}, 14\penalty0 (2):\penalty0
  195--219, 2012.

\bibitem[Ly et~al.(2015)Ly, Maggi, Montali, Rinderle-Ma, and van~der
  Aalst]{ly2015compliance}
L.~T. Ly, F.~M. Maggi, M.~Montali, S.~Rinderle-Ma, and W.~M. van~der Aalst.
\newblock {Compliance Monitoring in Business Processes: Functionalities,
  Application, and Tool-support}.
\newblock \emph{Information Systems}, 2015.

\bibitem[Martens and Teuteberg(2011)]{martens2011risk}
B.~Martens and F.~Teuteberg.
\newblock {Risk and Compliance Management for Cloud Computing Services:
  Designing a Reference Model}.
\newblock In \emph{AMCIS'11}, 2011.

\bibitem[Maxwell et~al.(2012)Maxwell, Anton, Swire,
  et~al.]{maxwell2012managing}
J.~C. Maxwell, A.~Anton, P.~Swire, et~al.
\newblock {Managing Changing Compliance Requirements by Predicting Regulatory
  Evolution}.
\newblock In \emph{20th Int. Conf. Requirements Engineering}, pages 101--110,
  2012.

\bibitem[Molina-Jimenez et~al.(2012)Molina-Jimenez, Shrivastava, and
  Strano]{molina2012model}
C.~Molina-Jimenez, S.~Shrivastava, and M.~Strano.
\newblock {A Model for Checking Contractual Compliance of Business
  Interactions}.
\newblock \emph{IEEE Transactions on Services Computing}, 5\penalty0
  (2):\penalty0 276--289, 2012.

\bibitem[Muller et~al.(2014)Muller, Oriol, Franch, Marco, Resinas,
  Ruiz-Cort{\'e}s, and Rodriguez]{muller2014comprehensive}
C.~Muller, M.~Oriol, X.~Franch, J.~Marco, M.~Resinas, A.~Ruiz-Cort{\'e}s, and
  M.~Rodriguez.
\newblock {Comprehensive Explanation of SLA Violations at Runtime}.
\newblock \emph{IEEE Transactions on Services Computing}, 7\penalty0
  (2):\penalty0 168--183, 2014.

\bibitem[Otto et~al.(2007)Otto, Ant{\'o}n, et~al.]{otto2007addressing}
P.~N. Otto, A.~Ant{\'o}n, et~al.
\newblock {Addressing Legal Requirements in Requirements Engineering}.
\newblock In \emph{15th Int. Conf. on Requirements Engineering}, pages 5--14,
  2007.

\bibitem[Panesar-Walawege et~al.(2013)Panesar-Walawege, Sabetzadeh, and
  Briand]{panesar2013supporting}
R.~K. Panesar-Walawege, M.~Sabetzadeh, and L.~Briand.
\newblock Supporting the verification of compliance to safety standards via
  model-driven engineering: Approach, tool-support and empirical validation.
\newblock \emph{Information and Software Technology}, 55\penalty0 (5):\penalty0
  836--864, 2013.

\bibitem[Ramezani et~al.(2012)Ramezani, Fahland, and van~der
  Aalst]{ramezani2012did}
E.~Ramezani, D.~Fahland, and W.~M. van~der Aalst.
\newblock {Where Did I Misbehave? Diagnostic Information in Compliance
  Checking}.
\newblock In \emph{10th Int. Conf. on Business Process Management}, pages
  262--278, 2012.

\bibitem[Ramezani et~al.(2014)Ramezani, Fahland, and van~der
  Aalst]{ramezani2014supporting}
E.~Ramezani, D.~Fahland, and W.~M. van~der Aalst.
\newblock {Supporting Domain Experts to Select and Configure Precise Compliance
  Rules}.
\newblock In \emph{Business Process Management Workshops}, pages 498--512,
  2014.

\bibitem[Schleicher et~al.(2009)Schleicher, Anstett, Leymann, and
  Mietzner]{schleicher2009maintaining}
D.~Schleicher, T.~Anstett, F.~Leymann, and R.~Mietzner.
\newblock {Maintaining Compliance in Customizable Process Models}.
\newblock In \emph{On the Move to Meaningful Internet Systems}, pages 60--75.
  Springer, 2009.

\bibitem[Schmidt et~al.(2012)Schmidt, Ant{\'o}n, Earp,
  et~al.]{schmidt2012assessing}
J.~Y. Schmidt, A.~Ant{\'o}n, J.~B. Earp, et~al.
\newblock {Assessing Identification of Compliance Requirements from Privacy
  Policies}.
\newblock In \emph{5th Int. Work. on Requirements Engineering and Law}, pages
  52--61, 2012.

\bibitem[Schumm et~al.(2010)Schumm, Leymann, Ma, Scheibler, and
  Strauch]{SchummLMS10}
D.~Schumm, F.~Leymann, Z.~Ma, T.~Scheibler, and S.~Strauch.
\newblock {Integrating Compliance into Business Processes: Process Fragments as
  Reusable Compliance Controls}.
\newblock In \emph{Multikonf. Wirtschaftsinformatik}, 2010.

\bibitem[Siena et~al.(2013)Siena, Ingolfo, Perini, Susi, and
  Mylopoulos]{siena2013automated}
A.~Siena, S.~Ingolfo, A.~Perini, A.~Susi, and J.~Mylopoulos.
\newblock {Automated Reasoning for Regulatory Compliance}.
\newblock In \emph{32nd Int. Conf. on Conceptual Modeling}, 2013.

\bibitem[{Tilburg University}(2008)]{Tilburg08}
{Tilburg University}.
\newblock {State-of-the-art in the Field of Compliance Languages}, 2008.

\bibitem[Tosatto et~al.(2015)Tosatto, Governatori, and
  Kelsen]{colombo2014business}
S.~Tosatto, G.~Governatori, and P.~Kelsen.
\newblock Business process regulatory compliance is hard.
\newblock \emph{IEEE Transactions on Services Computing}, 8\penalty0
  (6):\penalty0 958--970, 2015.

\bibitem[Toval et~al.(2002)Toval, Olmos, and Piattini]{toval2002legal}
A.~Toval, A.~Olmos, and M.~Piattini.
\newblock Legal requirements reuse: a critical success factor for requirements
  quality and personal data protection.
\newblock In \emph{Int. Conf. on Requirements Engineering}, pages 95--103,
  2002.

\bibitem[Tran et~al.(2012)Tran, Zdun, Oberortner, Mulo, Dustdar,
  et~al.]{tran2012compliance}
H.~Tran, U.~Zdun, E.~Oberortner, E.~Mulo, S.~Dustdar, et~al.
\newblock {Compliance in Service-oriented Architectures: a Model-driven and
  View-based Approach}.
\newblock \emph{Information and Software Technology}, 54\penalty0 (6):\penalty0
  531--552, 2012.

\bibitem[Turetken et~al.(2012)Turetken, Elgammal, Van~den Heuvel, and
  Papazoglou]{turetken2012capturing}
O.~Turetken, A.~Elgammal, W.-J. Van~den Heuvel, and M.~P. Papazoglou.
\newblock {Capturing Compliance Requirements: A Pattern-based Approach}.
\newblock \emph{Software, IEEE}, 29\penalty0 (3):\penalty0 28--36, 2012.

\bibitem[van~der Werf et~al.(2012)van~der Werf, Verbeek, and van~der
  Aalst]{van2012context}
J.~M.~E. van~der Werf, H.~Verbeek, and W.~M. van~der Aalst.
\newblock {Context-aware Compliance Checking}.
\newblock In \emph{10th Int. Conf. on Business Process Management}, 2012.

\bibitem[Zdun et~al.(2012)Zdun, Bener, and Olalia-Carin]{Zdun12IEEE}
U.~Zdun, A.~Bener, and E.~L. Olalia-Carin.
\newblock {Guest Editors' Introduction: Software Engineering for Compliance}.
\newblock \emph{Software, IEEE}, 29\penalty0 (3):\penalty0 24--27, 2012.

\end{thebibliography}
